\documentclass{ws-ijmpd}

\def\l{{\ell}}

\def\wmap{\hbox{\sl WMAP~}}

\def\etal{et al.~}
\def\alm{a_{\l m}}

\def\cl{C_{\l}}

\def\lm{\l m}

\def\G{{\bf G}}

\def\c{{\bf c}}

\def\d{{\bf d}}

\newcommand{\nbi}{{Niels Bohr Institute, Blegdamsvej 17,
DK-2100 Copenhagen, Denmark}}
\newcommand{\sao}{{Special Astrophysical Observatory, Nizhnij Arkhyz,
Karachaj-Cherkesia, 369167, Russia}}

\title{Peculiarities of phases of the WMAP quadrupole}
\author{Pavel D. Naselsky}
\address{\nbi, naselsky@nbi.dk}
\author{Oleg V. Verkhodanov}
\address{\sao, vo@sao.ru}

\date{Accepted 2006 November; Received 2006 September 29}

\begin{document}
\maketitle

\begin{abstract}
We present the analysis on the quadrupole phases of the Internal Linear
Combination map, ILC(I) and (III) derived by the \wmap team (1 and 3-year
data release). This approach allows us to see the global trend
of non-Gaussianity of the quadrupoles for the ILC(III) map through phase
correlations with the foregrounds.  Significant phase correlations is found
in between the ILC(III) quadrupole and the \wmap foregrounds phases for K-W
band: the phases of the ILC(III) quadrupole $\xi_{2,1}$, $\xi_{2,2}$ and
those of the foregrounds at K-W bands $\Phi_{2,1}$, $\Phi_{2,2}$ display
significant symmetry :  $\xi_{2,1}+\Phi_{2,1} \simeq \xi_{2,2}+\Phi_{2,2}$,
which is a strong indication that the morphology of the ILC(III) quadrupole
is mere reflection of that the foreground quadrupole through coupling.
To clarify this issue we exploit the symmetry of the CMB power, which is
invariant under permutation of the index $m=1\Leftrightarrow 2$. By simple
rotation of the ILC(III) phases with the same angle we reach the phases of
foreground quadrupole. We discuss possible sources of phase correlation and
come to the conclusion that the phases of the ILC(III) quadrupole reflect
most likely systematic effects such as changing of the gain factor for
the 3-year data release with respect to the 1-year, rather than
manifestation of the primordial non-Gaussianity.
\end{abstract}


\section{Introduction}
After the release of the Wilkinson Microwave
Anisotropy Probe (\wmap) 1-year results (hereafter WMAP\,I)
Refs.~\refcite{wmapresults},    % 1
\refcite{wmapfg},               % 2
\refcite{wmapsys},              % 3
\refcite{wmapcl}                % 4
the issue of non-Gaussianity of the CMB has attracted great attention.
In the papers
Refs.~\refcite{wmaptacng},      % 5
\refcite{park},                 % 6
\refcite{gonzalez},             % 7
\refcite{eriksen},              % 8
\refcite{hansen},               % 9
\refcite{larson},               % 10
\refcite{land},                 % 11
\refcite{roukema},              % 12
\refcite{vectors},              % 13
\refcite{chenalfven},           % 14
\refcite{McEwen},               % 15
various kinds of
methods have been employed and departure of Gaussianity have been detected.
These features can be of primordial origin
(Refs.~\refcite{eriksen},       % 8
\refcite{jaffe_2005})           % 16
or they could be related
to foreground residuals
(Refs.~\refcite{ndv04},         % 17
\refcite{autox},                % 18
\refcite{faraday})              % 19
or any sort of systematic effects
(\refcite{hansen}).             % 9

Recently the
\wmap team released the 3-year data (hereafter WMAP\,III)
Ref.~\refcite{wmap3ycos}        % 20
producing Internal  Linear Combination map (ILC(III))
available for scientific analysis at the multipole range
$\l\le 10$. The \wmap team performed the analysis of Gaussianity for the
ILC(III)
signal that does not include harmonics $\l \le 10$ and showed that
the CMB signal is Gaussian
Ref.~\refcite{wmap3ytem}.       % 21
However, as it was cautioned by the \wmap team, the alignment and
planarity at multipoles  $\l=2, 3$ and 5 mentioned in
Refs.~\refcite{eriksen},        % 8
\refcite{hansen},               % 9
\refcite{land},                 % 11
\refcite{roukema},              % 12
\refcite{schwarz}               % 22
still present in the map (see recent
investigation in
Ref.~\refcite{vectors1}).       % 23

If breaking of statistical isotropy in the low multipoles is of primordial
origin,
a very fundamental issue is consequently raised: what cosmological model
can provide such
a peculiar structure in the CMB angular distribution on the sky~?
In Refs.~\refcite{eriksen},     % 8
\refcite{jaffe_2006} and        % 24
\refcite{bridges}               % 25
it has been
pointed out
that breaking of statistical isotropy in the low multipoles can be
explained and in favor of
the Bianchi VIIh model. However,
Helling, Shupp and Tesilleanu (Ref.~\refcite{Helling}) % 26
have also pointed out
that $\l=2, 3$ and 5 have anti-alignment in the direction of non-cosmological
dipole and argued that accurate technique of the CMB signal subtraction
probably needs to be non-linear (with respect to dipole treatment)
to prevent any residuals from the dipole in the map.

In addition, methods of foreground separation for the CMB signal
very likely leave foreground residuals in the CMB map, a problem that
can be illustrated by cross-correlation between the derived signal and
the foregrounds. For the WMAP\,I data this problem was already investigated
in
Refs.~\refcite{toh},        % 27
\refcite{ndv03},            % 28
\refcite{ndv04},            % 17
\refcite{autox},            % 18
\refcite{n2},               % 29
\refcite{bielewicz}.        % 30

For WMAP\,III data, it has been shown in
Refs.~\refcite{chiang_t},   % 31
\refcite{autox}             % 18
that the low multipoles in ILC(III) clearly
display significant cross-correlation with the derived foregrounds,
while
Ref.~\refcite{cruz06}       % 32
showed that negative peak at $b=-57^{\circ}, l=209^{\circ}$) has no
analogue in the foreground map.
In
Refs.~\refcite{freeman},    % 33
\refcite{larson}            % 10
analyzing the
\wmap
map-making algorithm (MMA) it has been pointed out
that it could produce some residuals from the non-cosmological
dipole (related with the motion of our Galaxy in the Local Group) and any
non-Gaussian features can ``naturally'' arise as the result of the MMA.
Summarizing present status of non-Gaussianity and statistical anisotropy
of the CMB for low multipoles $l\le 10$, we may say that the origin of
these peculiarities still is  uncertain.

In this paper we go back to the question raised in
Refs.~\refcite{vectors}      % 13
and
\refcite{blandford},         % 34
if the \wmap quadrupole is cosmological~?
To answer
this question, we will implement the analysis of the ILC(I), ILC(III) and
de Oliveira-Costa \& Tegmark (Ref.~\refcite{doc_teg})  % 35
quadrupole and compare their phases
with those of the WMAP\,III
foregrounds\footnote{{\tt
http://lambda.gsfc.nasa.gov/product/map/m\_products.cfm}}:
combined maps from the synchrotron,
free-free, dust emission at K-W band. We will show that
combination of the ILC(III) phases $\xi_{2,m}$ and the foreground phases
$\Phi_{2,m}$ follow the intriguing relation
$\xi_{2,m}+\Phi_{2,m}\simeq const $
with accuracy, e.g. within $0.014$ (rad) for the foreground at K band.
Since phases of the signal is closely related to morphology of the
maps
(Ref.~\refcite{morphology}),                           % 36
detected
 correlation between the ILC(III) and foreground phases
allow us to conclude that it is most likely due to systematic effects
(such as estimation of the gain factor and foreground separation) rather then
primordial one.

\section{Phase analysis of the WMAP data}

For the statistical characterization of the CMB temperature anisotropies on
a sphere where $\theta$ and $\varphi$ are polar and azimuthal angle
of the polar system of coordinate, it is useful to express
$\Delta T(\theta,\varphi)$ in terms of spherical harmonics
$Y_{\lm}(\theta,\varphi)$:
\begin{equation}
\Delta T(\theta,\varphi)=\sum_{\l=0}^{\l_{\max}}\sum_{m=-\l}^\l |a_{\l,m}|
      \exp(i\phi_{\l,m}) Y_{\lm}(\theta,\varphi)
\label{eq1}
\end{equation}
where $|a_{\l,m}|$ and $\phi_{\l,m}$ are the amplitudes and phases
of the coefficients $a_{\l,m}$, respectively, and $|m|\le \l$.
$\Delta T(\theta,\varphi)$ can be full-sky signal, such as the \wmap
frequency maps from K to W band. We denote for the ILC and the foreground
coefficients as
$c_{\l,m}\equiv |c_{\l,m}|\exp(i\xi_{\l,m})$ and
$F^{(j)}_{\l,m}\equiv|F^{(j)}_{\l,m}|\exp(i\Phi^{(j)}_{\l,m})$, respectively. Here $\xi_{\l,m}$ and $\Phi^{(j)}_{\l,m}$ are the corresponding phases and the index $j=1-5$ marks the
\wmap frequency band for K, Ka, Q, V and W bands, respectively.
In standard cosmological models
(i.e. those involving the simplest forms of inflation) these temperature
fluctuations constitute a realization of a statistically homogeneous
and isotropic Gaussian random field. The statistical properties of
Gaussian random fields are completely determined by the power spectrum
\begin{equation}
\langle  a^{}_{\l^{ } m^{ }} a^{*}_{\l^{'} m^{'}} \rangle = \cl \;
\delta_{\l^{ } \l^{'}} \delta_{m^{} m^{'}},
\end{equation}
where the angle brackets indicate ensemble average. For any single
realization of signal, the angular power spectrum is estimated
\begin{equation}
\cl=\frac{1}{2\l+1} \sum_{m=-\l}^{\l} |\alm|^2.
\end{equation}
As the signal is always real, the conjugate properties of the spherical
harmonic coefficients allow us to write down the angular power spectrum as
\begin{equation}
C(\l)=\frac{1}{2\l+1}|c_{\l,0}|^2+\frac{2}{2\l+1}\sum_{m=1}^{\l}|c_{\l,m}|^2.
\label{pow}
\end{equation}
%xxxxxxxxxxxxxxxxxxxxxxxxxxxxxxxxxxxxxxxxxxxxxxxxxxxxxxxxxxxxxxxxxxxxxxxxxx

\subsection{Linear phase correlation method.}

To investigate different correlation of phases for different signals
(the CMB and foregrounds) we need to implement as much as possible
methods to detect corresponding correlations and then to show their
possible sources.
One of the possible method is to draw different linear combination
of the phases in form
\begin{eqnarray}
G_{l,m}=\sum_j\alpha_j\psi^{(j)}_{l,m}
\label{eQ}
\end{eqnarray}
where $\alpha_j$ is $(-1,0,1)$ and $\psi^{(j)}_{l,m}$ are the phases
of different signals $j$.
The basis of this method has simple motivation from the analysis of
the non-Gaussian random process recently performed in
Refs.~\refcite{cng}                      % 37
and
\refcite{mat}.                           % 38
If non-linearity of the random process $\delta(\vec{x})$, which leads
to its non-Gaussianity has simple quadratic form, the standard method
of the correlation analysis is the bispectrum.
It is defined as Fourier transform of the tree-point correlation function
\begin{eqnarray}
\zeta(\vec{r_1},\vec{r_2})=\langle\delta(\vec{x})\delta(\vec{x}+\vec{r_1})
\delta(\vec{x}+\vec{r_2})\rangle
\label{eQ1}
\end{eqnarray}
Generally speaking for any non-Gaussian processes we need to know all high
order cross-correlations or polyspectra, which are negligible for
the Gaussian random process.
However, we may characterize this sort of Non-Gaussianity in terms of
phase correlation.
For process characterized by the bispectrum only, corresponding equation
for the phase coupling, which maximizes the bispectrum, has a form
(Refs.~\refcite{cng}, \refcite{mat})           % 37, 38
\begin{eqnarray}
\G(\vec{k})=2\theta_{\vec{k}}-\theta_{\vec{2k}}; \hspace{0.5cm}
\G(\vec{k},\vec{p})=\theta_{\vec{k}} +\theta_{\vec{p}}-\theta_{\vec{p}+\vec{k}}
\label{eQ3}
\end{eqnarray}
where $\theta_{\vec{k}}$ is the Fourier phase of the random
process $\delta(\vec{x})$.

As one can see from Eq.(\ref{eQ3}), linear combination of the phases
characterizes non-linear coupling of the modes tested by bispectrum or
in general case by polyspectrum.
This is the reason why we introduce the functional $\G_{\l,m}$
to investigate different high order moments of the CMB-foreground
cross-correlation.

\subsection{Symmetry of the power spectrum.}
\begin{figure}
\centering
\epsfig{file=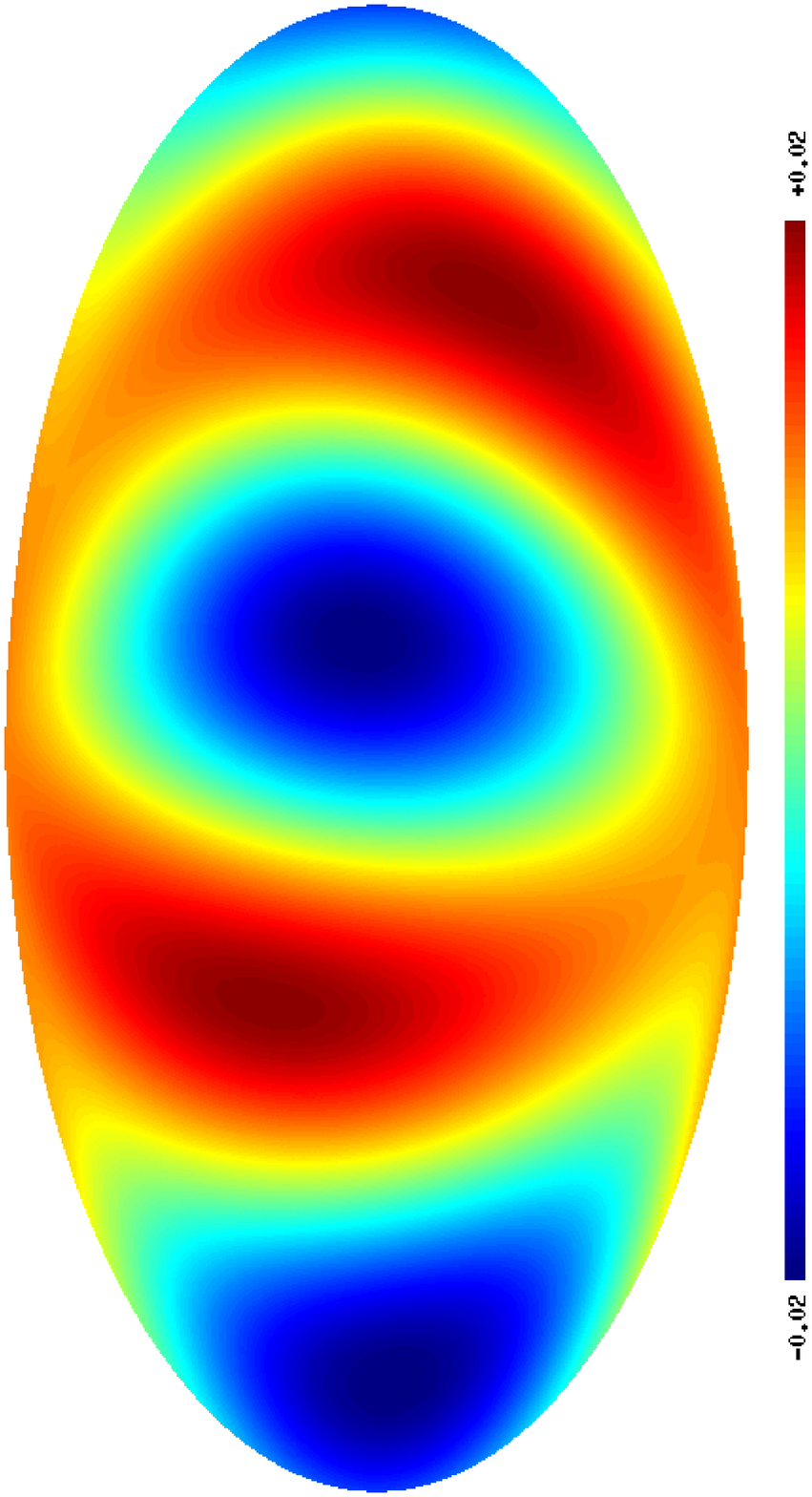 ,width=6cm}
\epsfig{file=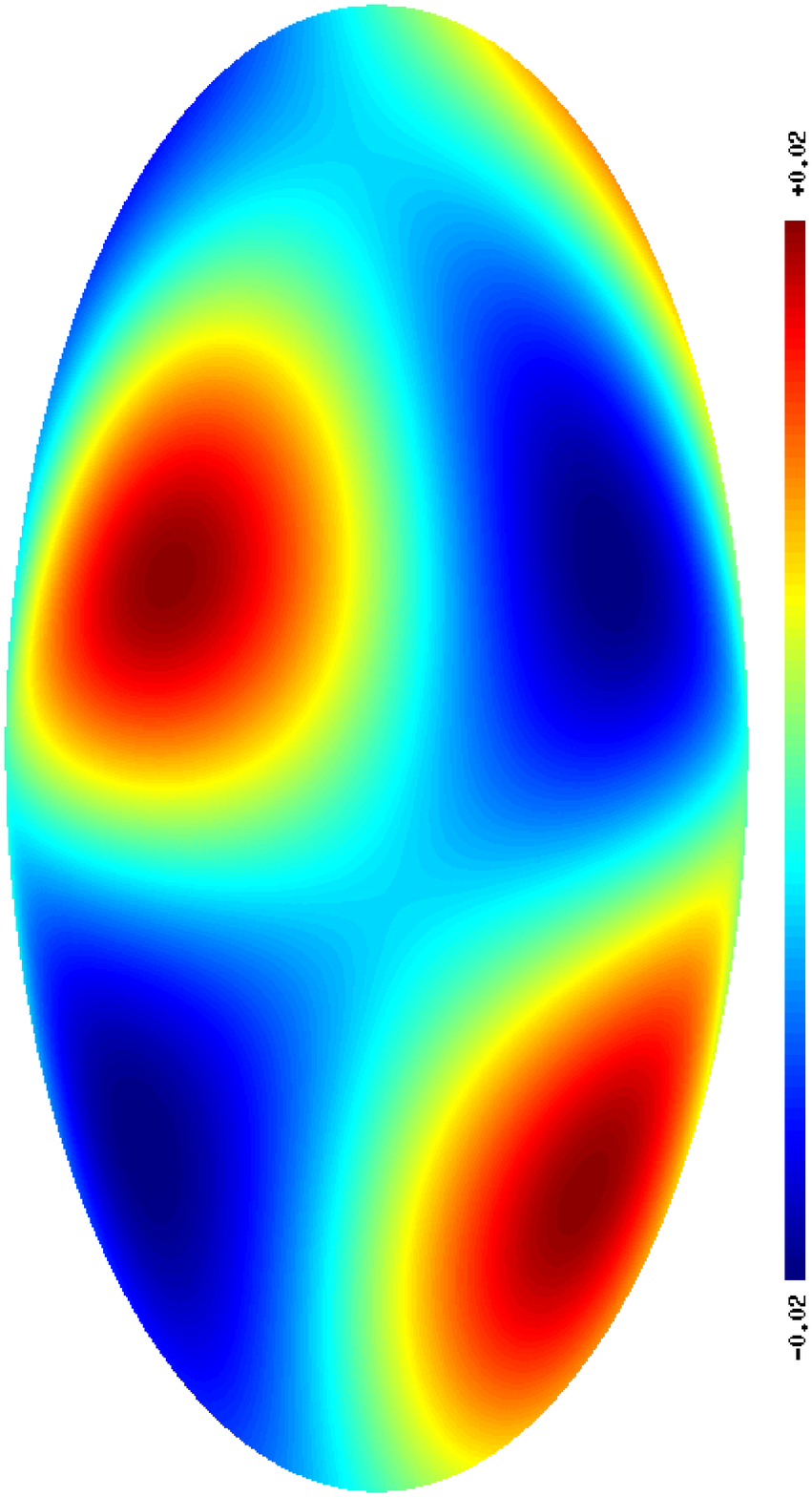 ,width=6cm}
\caption{The map for the ILC(III) quadrupole (top) and the transformed map
by the
permutation operator $\wp$ (bottom). These two maps have the same quadrupole
power.}
\label{figD}
\end{figure}

%xxxxxxxxxxxxxxxxxxxxxxxxxxxxxxxxxxxxxxxxxxxxxxxxxxxxxxxxxxxxxxxxxxxxxxxxxxxxxxxxxxxxxxxx
Let us discuss the properties of the second term in Eq.(\ref{pow}).
This part of the power spectrum is invariant under the following
transformation
$\d_{\l,m}=\wp \c_{\l,m^{'}}$, $d_{\l,m=0}=c_{\l,m=0}$, where $\wp$ is the
permutation operator for $m^{'}\le \l$. For the quadrupole component
\begin{eqnarray}
\left(
\begin{array}{r}
d_{2,1}\\
d_{2,2}
\end{array}
\right)
=\left(
\begin{array}{rr}
      ~ 0 ~ & ~ 1~  \\
      ~1~  & ~ 0~
\end{array}
\right)
\left(
\begin{array}{r}
c_{2,1}\\
c_{2,2}
\end{array}
\right),
\label{per}
\end{eqnarray}
i.e. this operator simply exchanges $m=1$ and 2 without altering
the power spectrum $C(\l)$. However, the map after such implementation of
the operator $\wp$
\begin{equation}
\Delta {\overline T(\theta,\varphi)}= \sum_{\l=2}^{\l_{\max}}\sum_{m=-\l}^\l|d_{\l,m}|e^{i\Psi_{\l,m}}Y_{\lm}(\theta,\varphi)
\end{equation}
is not invariant. As one can see from Eq.(\ref{per}),
the phases of the $d_{2,m}$
and $c_{2,m}$ coefficients correspond to the
following equations: $\Psi_{2,1}=\xi_{2,2}$ and  $\Psi_{2,2}=\xi_{2,1}$.
What is important is that if $c_{\l,m}$ constitute a Gaussian
random field the phases $\xi_{\l,m}$ should have no correlation
with the foreground phases $\Phi_{\l,m}$, which should also be the case for
the $\Psi_{\l,m}$ as well.
In Fig.\ref{figD} we plot the ILC(I) quadrupole in galactic
coordinates and the corresponding transformed map by the
permutation operator $\wp$, which have the same quadrupole power.
\section{Phase analysis of the WMAP first (WMAP\,I) and third (WMAP\,III)
years the CMB quadrupoles.}
\begin{figure}
\centering
\epsfig{file=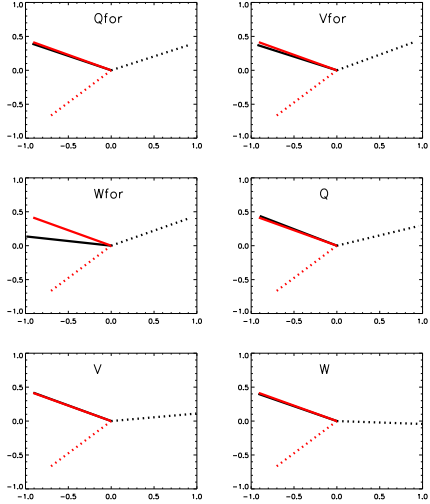 ,width=6cm}
\caption{Phases of the WMAP\,I Q, V and W band signals. The phase is
represented by the angle subtended between the positive $x$ axis and
the line. Solid lines are for $(\l,m)=(2,1)$ modes and dotted lines
for  $(\l,m)=(2,2)$. Red lines are the ILC(I) phases and are plotted
in all 6 panels whereas black lines are for either foreground maps or
frequency maps. Notice the overlapping of phases at the $(2,1)$ component
between the ILC(I) and Q and V foreground maps, and Q, V and W frequency
maps. The difference is listed in Table \ref{pdiffI}.}
\label{fig1}
\end{figure}

Below we examine the phases of the ILC and those of the foregrounds in more
detail.
Firstly, we plot the phases for the ILC(I) and the foreground and the
frequency maps from WMAP\,I in Fig.\ref{fig1}. The phases for foregrounds
at Q and V band and for frequency map at Q-W to show that
$\xi_{2,1} \simeq \Phi^{(3,4)}_{2,1}$ (see errors in the Table 1).
For K and Ka foregrounds the correlation are even stronger than for Q band.

As one can see from Table \ref{pdiffI}, the phase of the ILC(I) $\xi_{2,1}$
is extremely close to the phase of the foreground $\Phi_{2,1}$ while
the phase difference $\Phi_{2,2}-\xi_{2,2}$ is close to $\pi$.
To specify the probability of such realization of the phases
from uniformly distributed and non-correlated CMB phases with
the foregrounds,
we can use two vectors $\vec{n_1}=(\cos\xi_{2,1}, \sin\xi_{2,1})$
and $\vec{n_2}=(\cos\xi_{2,2}, \sin\xi_{2,2})$ for the ILC(I) and
$\vec{f_1}=(\cos\Phi_{2,1}, \sin\Phi_{2,1})$ and $\vec{f_2}=(\cos\Phi_{2,2},
\sin\Phi_{2,2})$ for the foregrounds.

%\begin{table}
\begin{table}
\centering
\begin{tabular}{|c|r|r|c|c|c|c|} \hline
 &$F^{\rm Q}$&$F^{\rm V}$&$F^{\rm W}$&Q&V&W \\ \hline
$(2,1)$&0.024 &0.044 &0.291 &-0.026 &-0.005  & 0.015  \\ \hline
$(2,2)$&3.508 &3.465 &3.479 &3.603  & 3.787  & 3.940  \\ \hline
\end{tabular}
\caption{The phase difference at $(\l,m)=(2,1)$ and $(2,2)$ mode between
the ILC(I) and the WMAP\,I foregrounds at Q, V and W band (left 3 columns)
and the WMAP\,I frequency bands (right 3 columns).}\label{pdiffI}
\end{table}

As is proposed for the multipole vector in
Refs.~\refcite{schwarz},    %  22
\refcite{Helling},          %  26
we can characterize the correlations of
these unit vectors in terms of scalar and vector product:
($\vec{n_{i}}\cdot\vec{f_{i}}$)
and ($\vec{n_{i}}\times\vec{f_{i}}$). Following Helling,
Shupp and Tesilleanu (Ref.~\refcite{Heeling}), we use the vector
product for estimation of the probability to get realization with
$\vec{n_{1}}\times\vec{f_{1}}=\sin(\Phi_{2,1}-\xi_{2,1})$ as
\begin{equation}
P(\Phi_{2,1},\xi_{2,1})=\frac{1}{2}C^2_1\sin2(\Phi_{2,1}-\xi_{2,1})
\end{equation}
where $C^k_p=k!/(k-p)!$. Since $\Phi_{2,1}-\xi_{2,1}\ll \pi/2 $,
this probability corresponds to the phase difference
$\Phi_{2,1}-\xi_{2,1} $ multiplied by a factor of 2.
From Table 1 one can see
that for Q band foreground it is about 5\% .

\begin{figure}
\centering
\epsfig{file=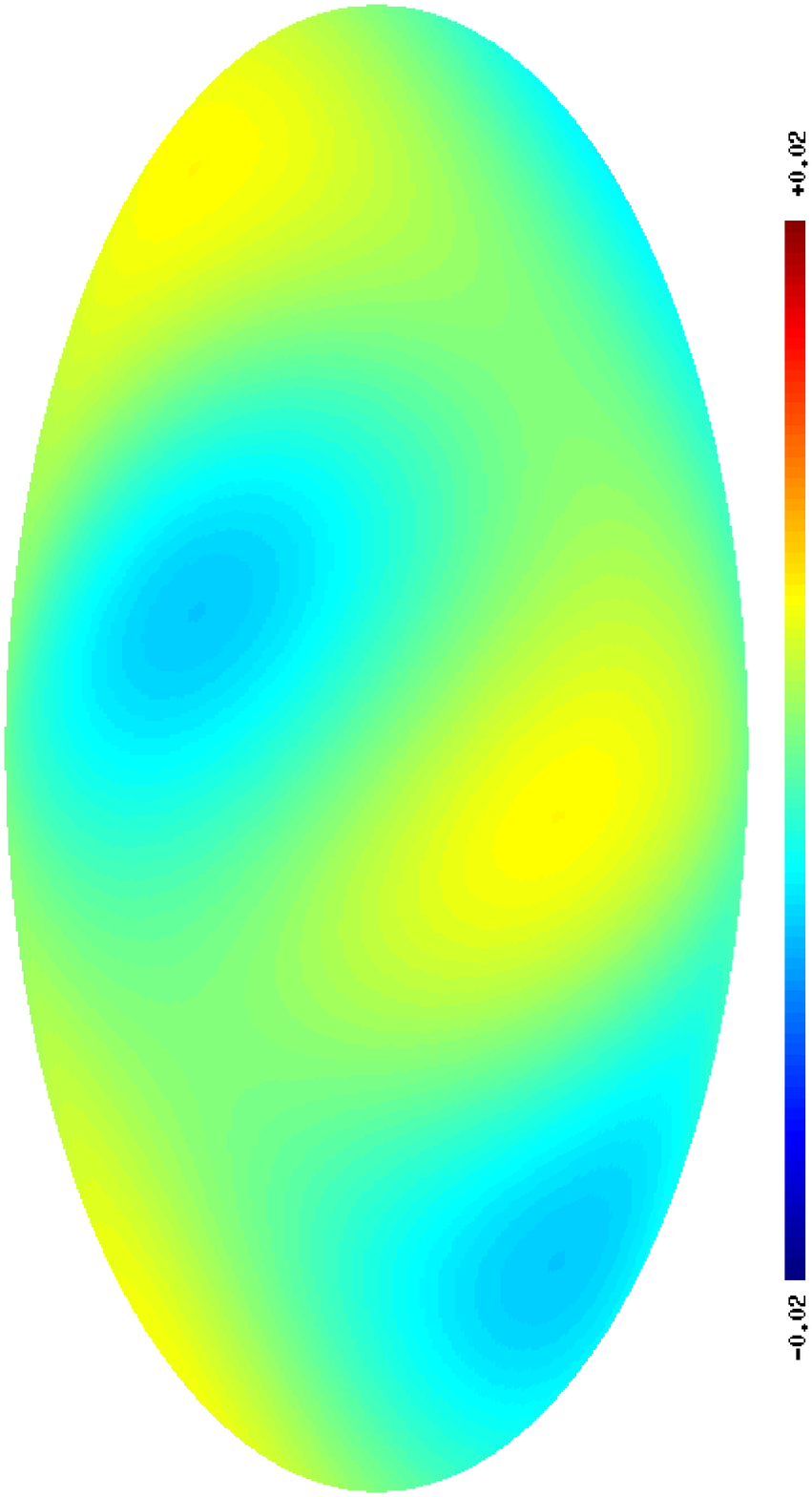 ,width=6cm}
\epsfig{file=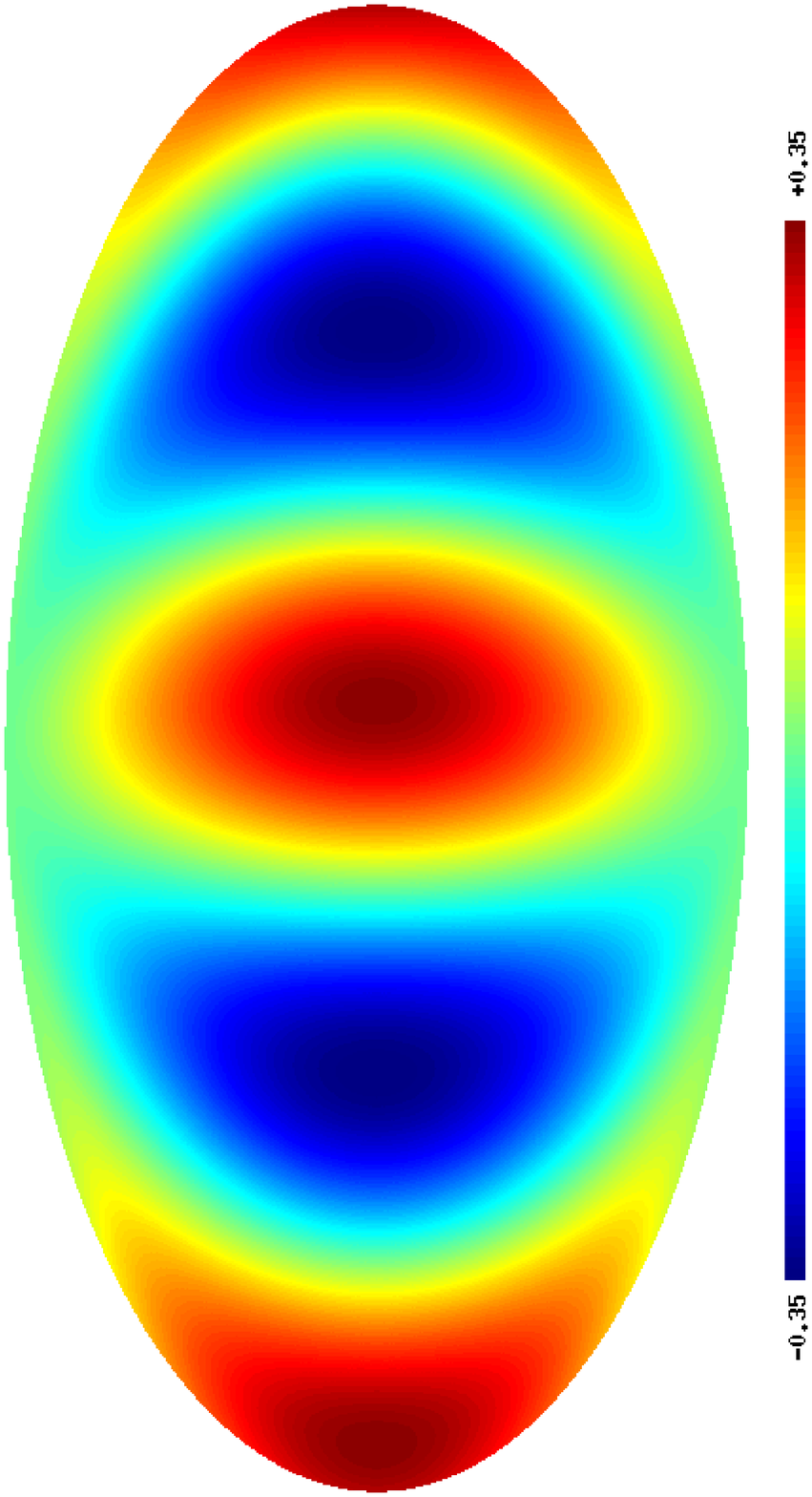,width=6cm}
\caption{The map for difference between the ILC(III)and ILC(I) (top)
and that for difference between
the K band foreground WMAP\,III and WMAP\,I (bottom).}
\label{dif}
\end{figure}

As is claimed by the \wmap science team, improvement
of systematic effects, including the gain factor, leads to correction
of the foregrounds and the ILC(I) map. We show below that this
improvement reveals significant correlations between the ILC(III),
the foregrounds and non-cosmological dipole phases.
First, we start from analysis of the difference between ILC(III)
and ILC(I) map. In Fig.\ref{dif} we plot the quadrupole map
for ILC(III)-ILC(I) in Galactic coordinates.

\begin{figure}
\centering
\epsfig{file=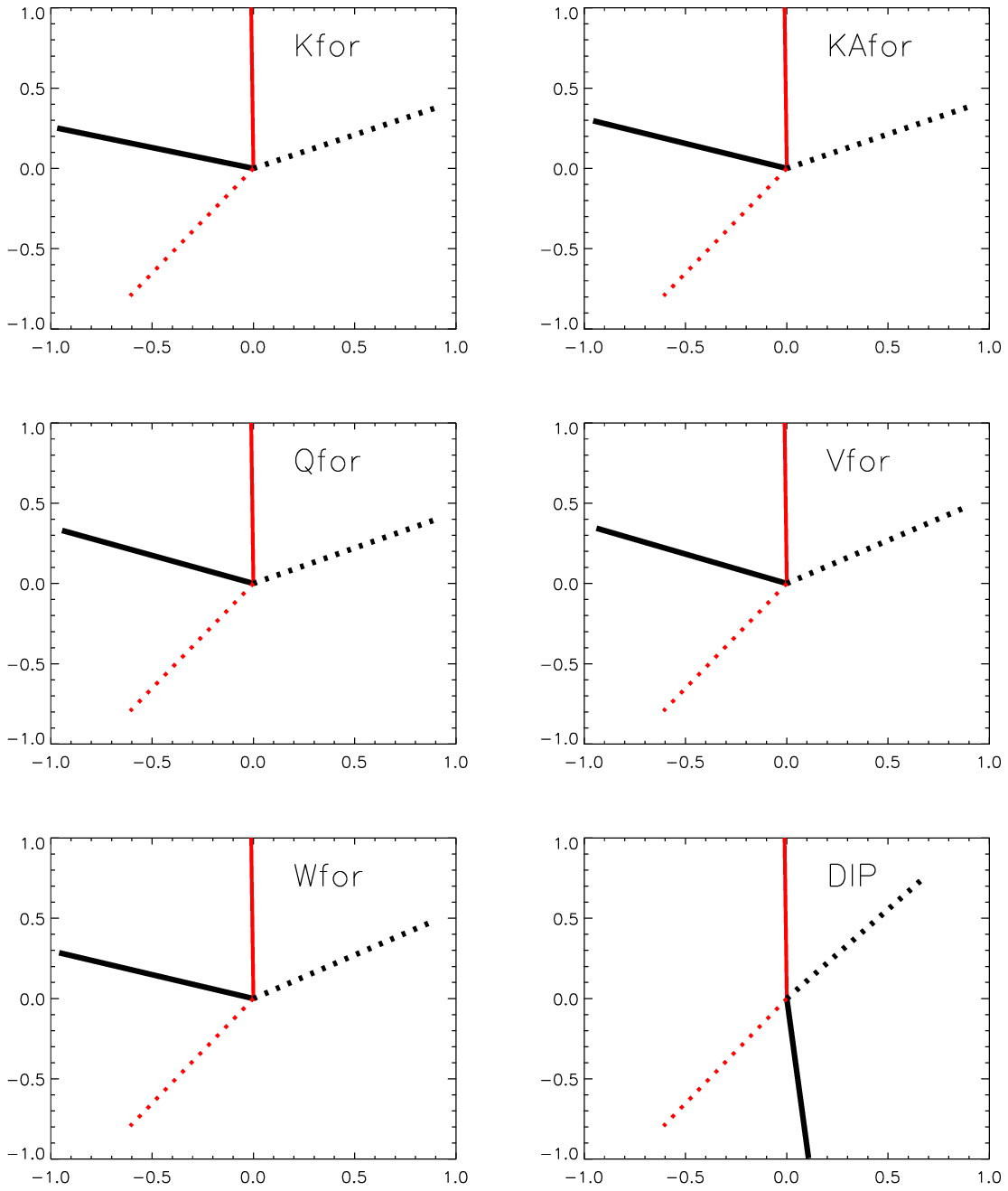 ,width=6cm}
\caption{Phases of WMAP\,III foregrounds and the ILC(III). The phase
is represented by the angle subtended between the positive $x$ axis
and the line. Solid lines are for $(\l,m)=(2,1)$ modes and dotted lines
for  $(\l,m)=(2,2)$. Black lines are for foregrounds and red lines are
for ILC(III). The bottom right panel shows  the ILC(III) phases in red,
the non-cosmological dipole ($(\l,m)=(1,1)$) phase (black solid line) and
the  cosmological dipole one (black dotted line).  }
\label{fig0}
\end{figure}

Correction of the foregrounds in WMAP\,III and the ILC(III) in comparison
with
WMAP\,I leads to the changes of the ILC(III) and the foregrounds phases as
shown in Fig.\ref{fig0}.
In Table \ref{pdiffIII} we list the phase differences for
$\Phi_{2,1}-\Phi_{2,2}$
and $\xi_{2,2}-\xi_{2,1}$. As one can see from Fig\ref{fig0},
the phases of ILC(III) now are
significantly different with respect to the ILC(I), while the phases of
quadrupole for the foregrounds at $(2,2)$ are still shifted
by $\sim \pi$. The bottom right panel marked with ``DIP'' in Fig\ref{fig0}
shows the shift of
phases for the ILC(III) quadrupole with respect to both non-cosmological
and cosmological dipole phase ($(\l,m)=(1,1)$) from ILC(III).

From K and Ka band of the Table \ref{pdiffIII} one
can write down an empirical equation
\begin{eqnarray}
\xi_{2,1}+\Phi_{2,1}&\simeq &\xi_{2,2}+\Phi_{2,2}= 4.46963 (-0.0147)\hspace{0.25cm};{\rm K} \nonumber\\
\xi_{2,1}+\Phi_{2,1}&\simeq &\xi_{2,2}+\Phi_{2,2}= 4.42253 (+0.0396)\hspace{0.25cm};{\rm Ka} \nonumber\\
\label{vic}
\end{eqnarray}

\begin{table}
\centering
\begin{tabular}{|r|r|c|c|c|c|}
\hline
 &$F^{\rm K}$&$F^{\rm Ka}$&$F^{\rm Q}$&$F^{\rm V}$&$F^{\rm W}$ \\ \hline
$\Phi^{(i)}_{2,1}-\Phi^{(i)}_{2,2}$&2.4916&2.4373&2.3886&2.2974&2.3557\\ \hline
$(\xi_{2,2}-\xi_{2,1})-$& & & & & \\
$(\Phi^{(i)}_{2,1}-\Phi^{(i)}_{2,2})$&0.0147&0.0396&0.0881&0.2471&0.1211\\ \hline
\end{tabular}
\caption{Phase differences between $(2,1)$ and $(2,2)$ for WMAP\,III K to W band foregrounds (top row) and differences between the difference of ILC(III) and those of the foregrounds (bottom row).}\label{pdiffIII}
\end{table}
The probability for the phases of the foregrounds and
the non-correlated phases to satisfy Eq.(\ref{vic}) is exactly
the modulo of difference between $\xi_{2,1}+\Phi_{2,1}$ and
$\xi_{2,2}+\Phi_{2,2}$, as shown in the brackets in Eq.(\ref{vic}).

Correlation of phases in the form of Eq.(\ref{vic}) can be easily explained
in terms of phase correlation after the permutation operation on the phases.
Taking account of $\Psi_{2,1}=\xi_{2,2}$ and $\Psi_{2,2}=\xi_{2,1}$,
from Eq.(\ref{vic}) we obtain $\Psi_{2,1}-\Phi_{2,1} \simeq
\Psi_{2,2}-\Phi_{2,2}$ and rotation of the $d_{2,m}$ phases
by the angle $\Delta=\Psi_{2,1}-\Phi_{2,1}$ transforms
the phases $\Psi_{2,1}-\Delta,\Psi_{2,2}-\Delta$  to the phases of the
foregrounds $\Phi_{2,1},\Phi_{2,2}$.

As is shown in Table \ref{pdiffIII}, the probability for ILC and
foreground phases
to follow Eq.(\ref{vic}) for uniformly
distributed and statistically independent from the foregrounds
Gaussian CMB signal increase from $\sim 0.09$ for Q band and reach
the maximum 0.22 for the V band and falls down to
$0.12$ for the W band. This result seems to have natural explanation,
since the contribution of the synchrotron emission to the Q, V and W bands
decrease, and the free-free and dust emission dominate over synchrotron
emission.

\section{Multipole vectors analysis}

In this section we compare the phase correlation method with multipole
vectors approach, proposed in
Refs.~\refcite{vectors},         %  13
\refcite{vectors1}.              %  23
We apply the multipole vectors  method for the
ILC(III) quadrupole and  kinetic quadrupole.
After this we calculate the  multipole vectors
using the method in Ref.~\refcite{vectors1}.

The values of ILC quadrupole $\alm$ are:
\begin{eqnarray}
\begin{array}{cccc}
\l   &  m   &    \Re           &     \Im           \\ \nonumber
2      &   0  & +1.147576980\times10^{-2} & 0  \\ \nonumber
2      &   1  & -5.329925261\times10^{-5} & +4.864335060\times10^{-3}  \\ \nonumber
2      &   2  & -1.440715604\times10^{-2} & -1.880360022\times10^{-2} \nonumber
\end{array}
\end{eqnarray}
After correction of the kinetic quadrupole, we have
\begin{eqnarray}
\begin{array}{cccc}
\l   &  m &    \Re              &     \Im           \\ \nonumber
2      &   0  & +1.147431601\times10^{-2} &  0 \\ \nonumber
2      &   1  & -5.358225098\times10^{-5} & +4.866961855\times10^{-3} \\ \nonumber
2      &   2  & -1.440600399\times10^{-2} & -1.880334876\times10^{-2} \nonumber
\end{array}
\end{eqnarray}

According to
Ref.~\refcite{vectors}          % 13
we describe these $a_{\l m}$
by multipole vector with coordinates
$(-0.562380$,  $0.815276, 0.138036)$ and
$(0.970920,  0.048491, 0.234440)$
which correspond to the points on a sphere with Galactic coordinates (in degrees)
$(l=124.6, b=  7.9)$ and $(l=  2.9, b= 13.6)$ respectively.

We perform the same conversion for the K band. The input data are
\begin{eqnarray}
\begin{array}{cccc}
\l &  m &    \Re            &     \Im           \\ \nonumber
2      &   0  & -2.167431593\times10^{+0}  & 0   \\ \nonumber
2      &   1  & -1.421663761\times10^{-1} & 4.126564786\times10^{-2}   \\ \nonumber
2      &   2  & +4.031517506\times10^{-1} & 1.558263451\times10^{-1} \nonumber
\end{array}
\end{eqnarray}
After correction for the kinetic quadrupole, we have
\begin{eqnarray}
\begin{array}{cccc}
\l &  m &    \Re            &     \Im            \\ \nonumber
2      &   0  &   -2.167433023\times10^{+0}   & 0  \\ \nonumber
2      &   1  &   -1.421666592\times10^{-1} & 4.126827419\times10^{-2}  \\ \nonumber
2      &   2  &   +4.031529129\times10^{-1} & 1.558265984\times10^{-1} \nonumber
\end{array}
\end{eqnarray}
This gives us the corresponding multipole vector coordinates and
the Galactic positions on a sphere:
$(-0.560275,  0.076989,  0.824721)$, $(l=172.2, b= 55.6)$
and
$(-0.479820,  0.113498, -0.869995)$, $(l=166.7, b=-60.5)$.

Using statistics for such vectors, we have found area vectors in a form
\begin{equation}
  w^{(2;1,2)}  = v^{(2,1)} \times v^{(2,2)}.
\end{equation}
This gives us the area vectors with coordinates
$w_{III}$: $(0.1850,0.2657,-0.2194)$ or $(l=55.2, b=-34.1)$
and
$w_{K}$: $(-0.1602,-0.8826,0.0033)$ or $(l=259.7, b=0.2)$.

After this we calculated statistics following Ref.~\refcite{vectors1}
to estimate correlation properties:
dot product of ILC and K-channel vectors
\begin{equation}
   d^{(2,i)} = v^{(2,i)}_{III} \cdot v^{(2,i)}_K \,,
\end{equation}
dot product of ILC and K-channel area vectors
\begin{equation}
   \Delta^{(2;1,2)} = \frac{w^{(2;1,2)}_{III}\cdot w^{(2;1,2)}_{K}}
	{ |w^{(2;1,2)}_{III}| |w^{(2;1,2)}_{K}|} \,,
\end{equation}
and ratio of lengths
\begin{equation}
   r^{(2;1,2)} = \frac {|w^{(2;1,2)}_{K}|} {|w^{(2;1,2)}_{III}|}
\end{equation}

So, we have got
\begin{eqnarray}
\begin{array}{ll}
\centering
 d^{(2,1)}         & +0.4908 \\ \nonumber
 d^{(2,2)}         & -0.6643 \\ \nonumber
 \Delta^{(2;1,2)}  & -0.7550 \\ \nonumber
 r^{(2;1,2)}       & +2.2939
\end{array}
\label{M}
\end {eqnarray}
As one can see from this Table, all the dot and cross-products are not
especially close to unity (see for comparison Table 2 in
Ref.~\refcite{vectors1}               % 23
for ILC(I) and ILC(III), for which
$d^{(2,1)}\simeq 0.973$,$d^{(2,2)}\simeq 0.956$,
$\Delta^{(2;1,2)}\simeq 0.955$ and $r^{(2;1,2)}\simeq 0.952$),
which means that the multipole vector approach is not very sensitive
for analysis of the CMB-foreground correlations.

\section{Discussions}

\subsection{Systematic effects~?}
For interpretation of correlations between the \wmap low
multipoles (i.e. alignment for $\l=2,3$, planarity for $\l=5$ etc.) we need
to know how possibly systematic effects determine their properties.
This problem will be even more crucial if one is to explain the deficit
of the ILC quadrupole power. We will need to implement
modification of the theory of inflation (cut-off in power
for primordial adiabatic fluctuations at spatial scales $>10^3$ Mpc),
the Bianchi VIIh model, or any theoretical changes of the cosmological
scenario, assuming primordial origin of the ILC(III) low amplitude
quadrupole.
\begin{figure}
\centering
\epsfig{file=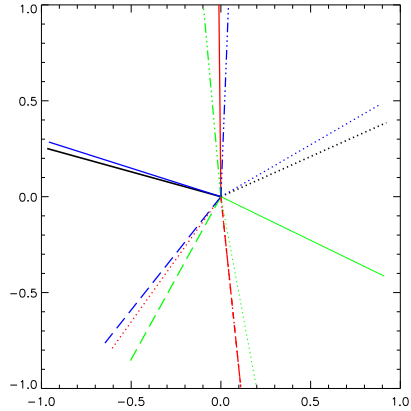,width=6cm}
\caption{Phases of WMAP\,III foregrounds and the ILC(III).
The phase is represented by the angle subtended between the
positive $x$ axis and the line. Solid lines are for $(\l,m)=(2,1)$
modes and dotted lines for  $(\l,m)=(2,2)$. Red lines are for ILC(III),
black and blue lines for K and W band foregrounds, respectively. Green
for ILC(III)-ILC(I). Thick red dash line is for non-cosmological dipole.
Thick blue and green are for M05 and M06 respectively, in which dash
lines and triple dot-dash lines are for $(2,1)$ and $(2,2)$, respectively.}
\label{figA}
\end{figure}

Probably, the best and more significant illustration of the systematic
effects is the improvement of the gain factor performed by the
\wmap science team, which could be the reason why the quadrupole
from ILC(I) has
different amplitude and phases in comparison with from ILC(III).
More importantly, the quadrupole power of the ILC(I) and the ILC(III)
does not change as much as the phases reveal of a new
sort of correlation between the ILC(III) phases and the foreground
phases (see Eq.(\ref{vic})). In Fig.\ref{figA} we plot the corresponding
phases for the map of difference between the ILC(III) and the ILC(I)
quadrupoles.
Next type of systematic effect is illustrated by the comparison of
the ILC(III) quadrupole and the quadrupole derived by
de Oliveira-Costa and Tegmark (Ref.~\refcite{doc_teg})
from the WMAP\,I and the WMAP\,III.
Difference
between the ILC(III) and
de Oliveira-Costa \& Tegmark (Ref.~\refcite{doc_teg})
quadrupoles is mainly related with different Galactic mask, which we
will call M05 and M06 (following Ref.~\refcite{doc_teg}).
As one can see from Fig.\ref{figA} and Table 3, the error of
the $\xi_{2,1}$ ILC phase reconstruction is about $0.1$ radians for M06
and it is negligible for M05 mask. For $\xi_{2,1}$ the corresponding error
is $\sim 0.12$ radians (M06) and $\sim 0.05$ radians (M05).
This means that with
uncertainties about $\pm 0.12$ radians Eq.(\ref{vic}) is correct
for all the K-W foregrounds.
\begin{table}
\label{temperature}
\begin{center}
\begin{tabular}{|r|r|c|c|c|}
\hline
    &ILC(III)&$\Delta$&M05& M06 \\
 \hline
$(2,1)$&1.582 & 5.856&1.528 &1.668  \\
\hline
$(2,2)$&4.059 &4.909 & 4.010&4.181\\
 \hline
\end{tabular}
\end{center}
\small
\caption{The table of phase difference for K-W foreground and
the ILC(III) for Galactic coordinates. For the foregrounds we use
the phase difference $\Phi_{2,1}-\Phi_{2,2}$, while for ILC(III)
we use  $\xi_{2,2}-\xi_{2,1}$.}
\end{table}

Let us discuss another important
correlations of the ILC(III) phases and the phases of non-cosmological
and ``cosmological'' dipoles shown in Fig.\ref{fig0}. By dipole phase we mean the phase of $(\l,m)=(1,1)$ mode. The angle between $\xi_{2,1}=1.528$ and the non-cosmological dipole $\Psi_{nc}=4.824$
is about $\alpha=\Psi_{nc}-\xi_{2,1}=3.295$ radians, while the angle between
$\xi_{2,2}=4.0586$ and the phase of cosmological dipole $\Psi_c=0.8379$
is $\beta=\xi_{2,2}-\Psi_c=3.2207$.
Thus, $\alpha-\beta\simeq 0.074$ rad for the ILC(III).

We would like to point out that the above-mentioned correlations between
the phases of the ILC(III) and those of the WMAP\,III foregrounds are
mainly related to the changes of the gain factor. The crucial part of
these correlations is the angle between
$\xi_{2,2}-\xi_{2,1}$ and for the foreground phases $\Phi_{2,2}-\Phi_{2,1}$.
Note that transition from the WMAP\,I to WMAP\,III leads to rotation of
the phases for the ILC and for
the foregrounds as well. Since WMAP\,III data seem more accurate
in terms of systematic effect removal, decreasing of direct
correlations between $\xi_{2,1}$ and $\Phi_{2,1}$, typical for
the WMAP\,I, reveals another type of correlation
between the ILC(III) and the foregrounds. In particular, in addition
to Eq.(\ref{vic}) one can get
\begin{eqnarray}
\xi^{\rm I}_{2,2}-\xi^{\rm I}_{2,1}\simeq 1.186 ~rad.\nonumber\\
\xi^{\rm I}_{2,1}-\xi^{\rm III}_{2,1}\simeq 1.133~rad.
 \label{ww}
\end{eqnarray}
which means that the phases for the ILC(I) and ILC(III) have some regular
changes. It is quite possible that these changes reflect some
correlations between the ILC(I) phases and the WMAP\,I foregrounds.
The comparison between synchrotron phases for the K band shows that
from Eq.(\ref{ww}) one can obtain
$\xi^{\rm I}_{2,2}-\xi^{\rm III}_{2,1}\simeq 2.31907$ radians,
while $\Phi^{\rm I}_{2,1}-\Phi^{I}_{2,2}\simeq 2.34936$ radians.
So, with accuracy about $0.030$ radian we get :
$\xi^{\rm III}_{2,1}\simeq \xi^{\rm I}_{2,2}-(\Phi^{\rm I}_{2,1}-\Phi^{\rm I}_{2,2})$.
On the other hand, from Eq.(\ref{ww}) one can get
$\xi^{\rm III}_{2,1}\simeq 2\xi^{\rm I}_{2,1}-\xi^{\rm I}_{2,2}$.
Finally, combining these
two equations we get:
\begin{equation}
2\xi^{\rm I}_{2,1}+\Phi^{\rm I}_{2,1}\simeq 2\xi^{\rm I}_{2,2}+\Phi^{\rm I}_{2,2}
\label{ww1}
\end{equation}
As for the WMAP\,III, one can see that the phases of the ILC(I)
and the WMAP\,I foregrounds per K band are strongly
coupled in the form of Eq.(\ref{ww1}). However, note that
the Eq.(\ref{ww1}), like the Eq.(\ref{vic}), is  ``empirical'',
displaying possible relationship
between the phases of the ILC and the foregrounds, which requires explanation
in the framework of the CMB and foreground separation scheme.

Implemented in this section phase approach can be reformulate in terms
of standard correlation between different components of quadrupoles.
Since the quadrupole component of the signals (the CMB and the foregrounds)
contain only $a_{2,1}$ and $a_{2,2}$, the coefficient of
the cross-correlation between them is simply
\begin{equation}
K^{\rm ILC}_{1,2}=\frac{|a_{2,1}||a_{2,2}|
     \cos(\xi_{2,1}-\xi_{2,2})}{\sqrt{|a_{2,1}|^2|a_{2,2}|^2}} =
	  \cos(\xi_{2,1}-\xi_{2,2})
\end{equation}
Even if the ILC quadrupole is of Gaussian nature, for a single realization of
the random process, as our Universe does, this cross-correlation
coefficients are not informative apart from some peculiar values
such as 0 or 1.
However, combining the coefficient $K^{\rm ILC}_{1,2}$ with
the same coefficient $K^{f}_{1,2}$,
defined for the WMAP foregrounds by simple comparison between these
two numbers we can conclude about common morphology of these signals.

\subsection{Primordial origin ?}

\begin{figure}
\centering
\epsfig{file=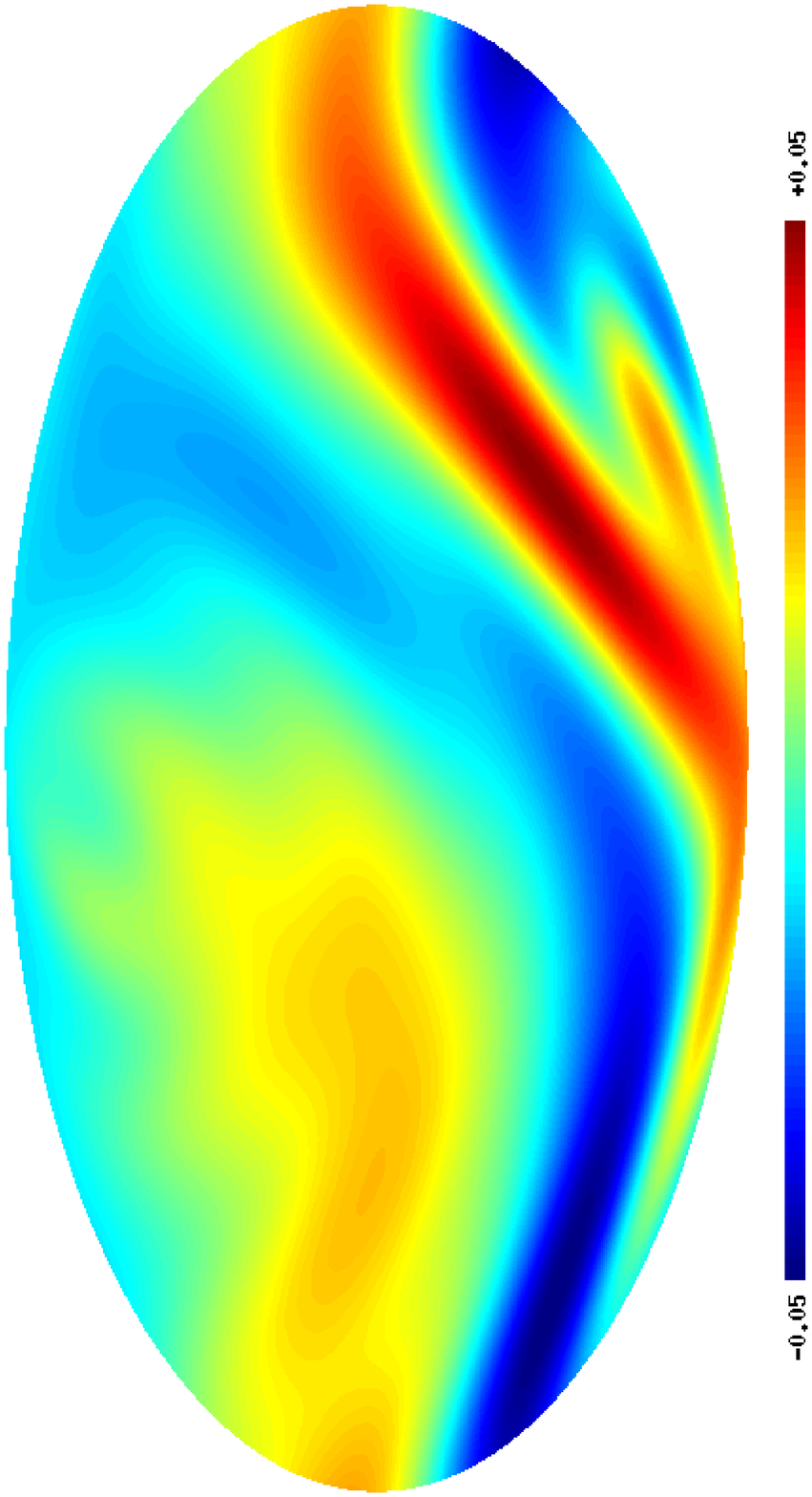,width=6cm}
\epsfig{file=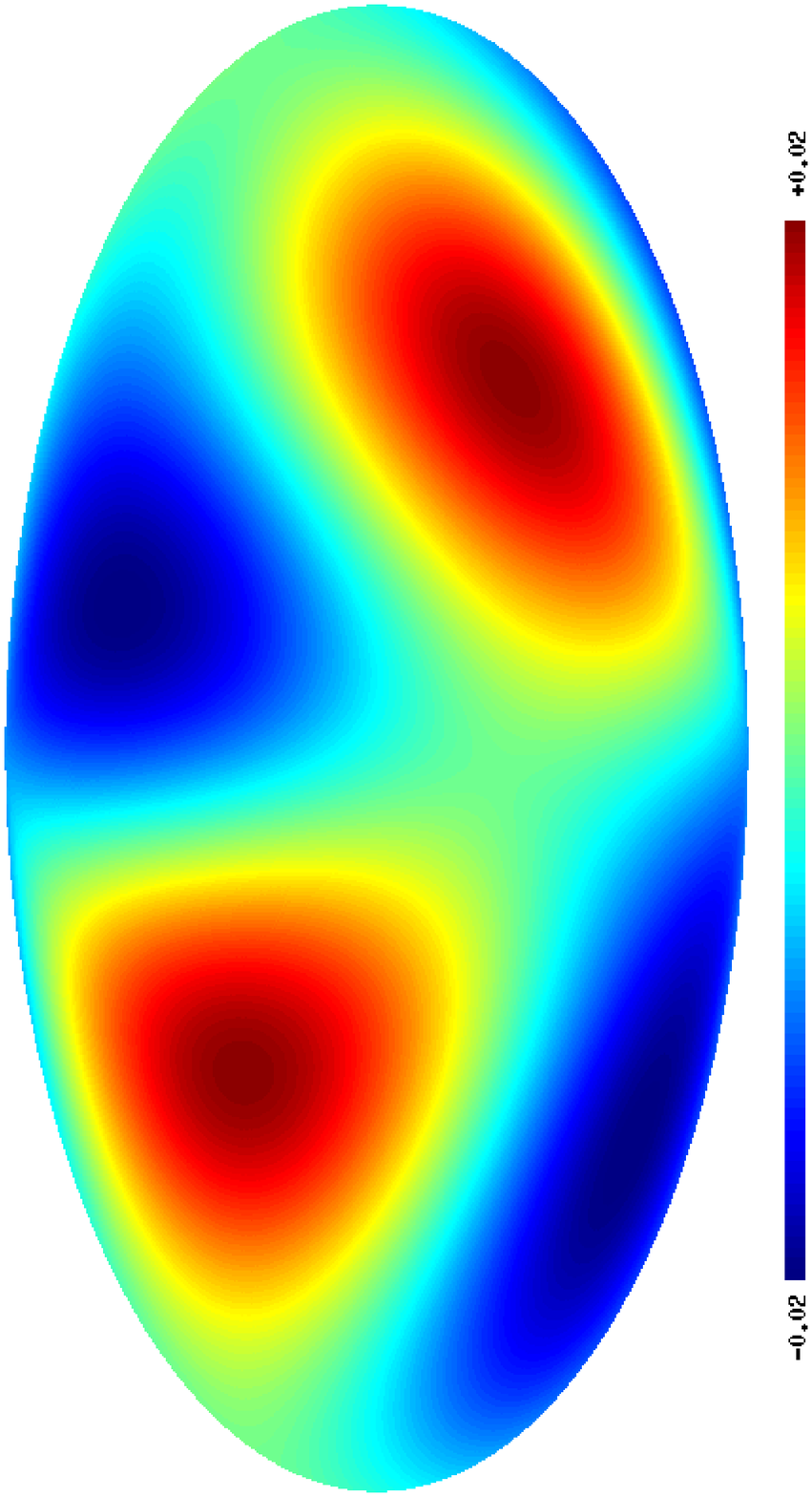,width=6cm}
\epsfig{file=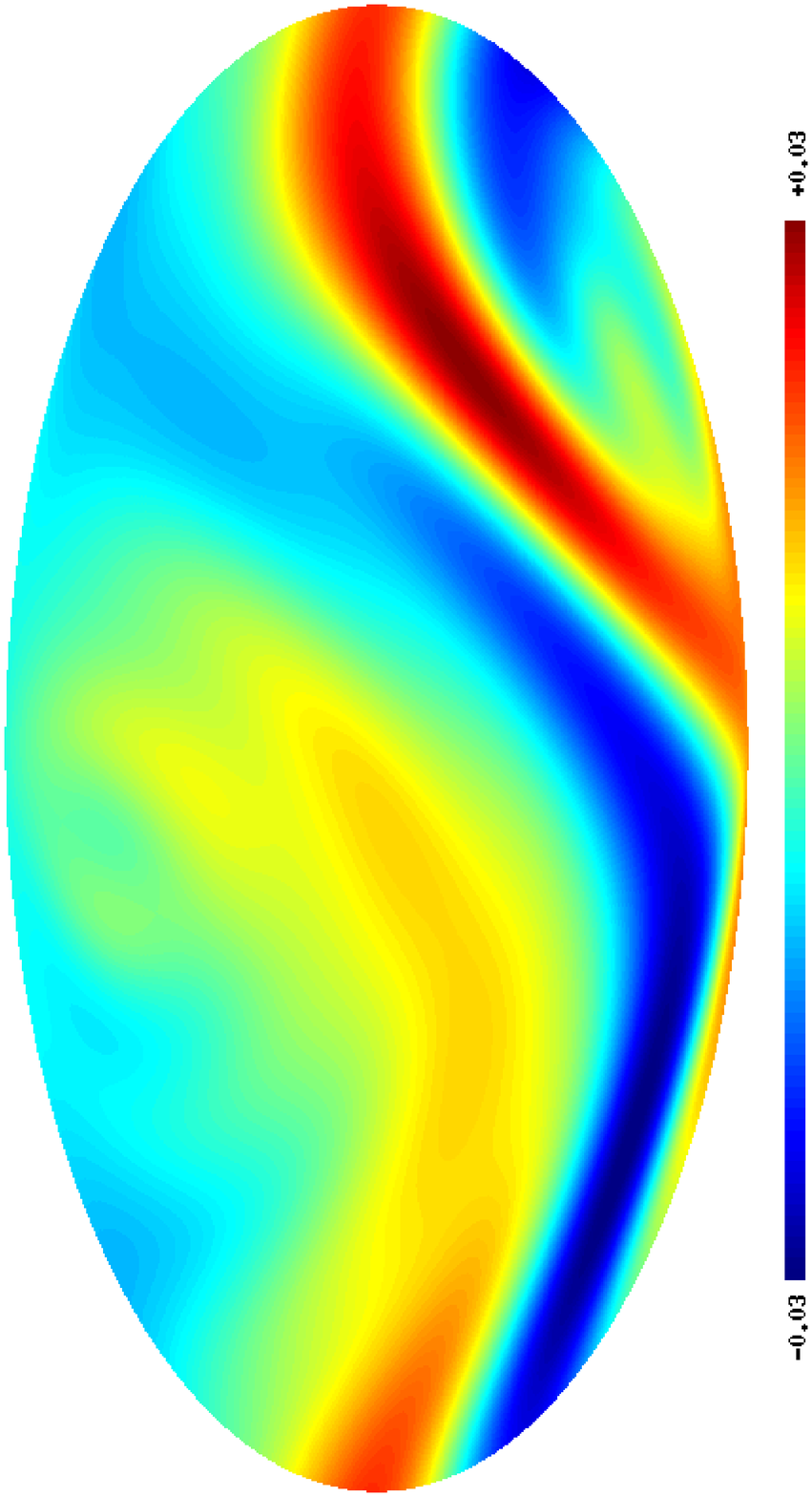,width=6cm}
\epsfig{file=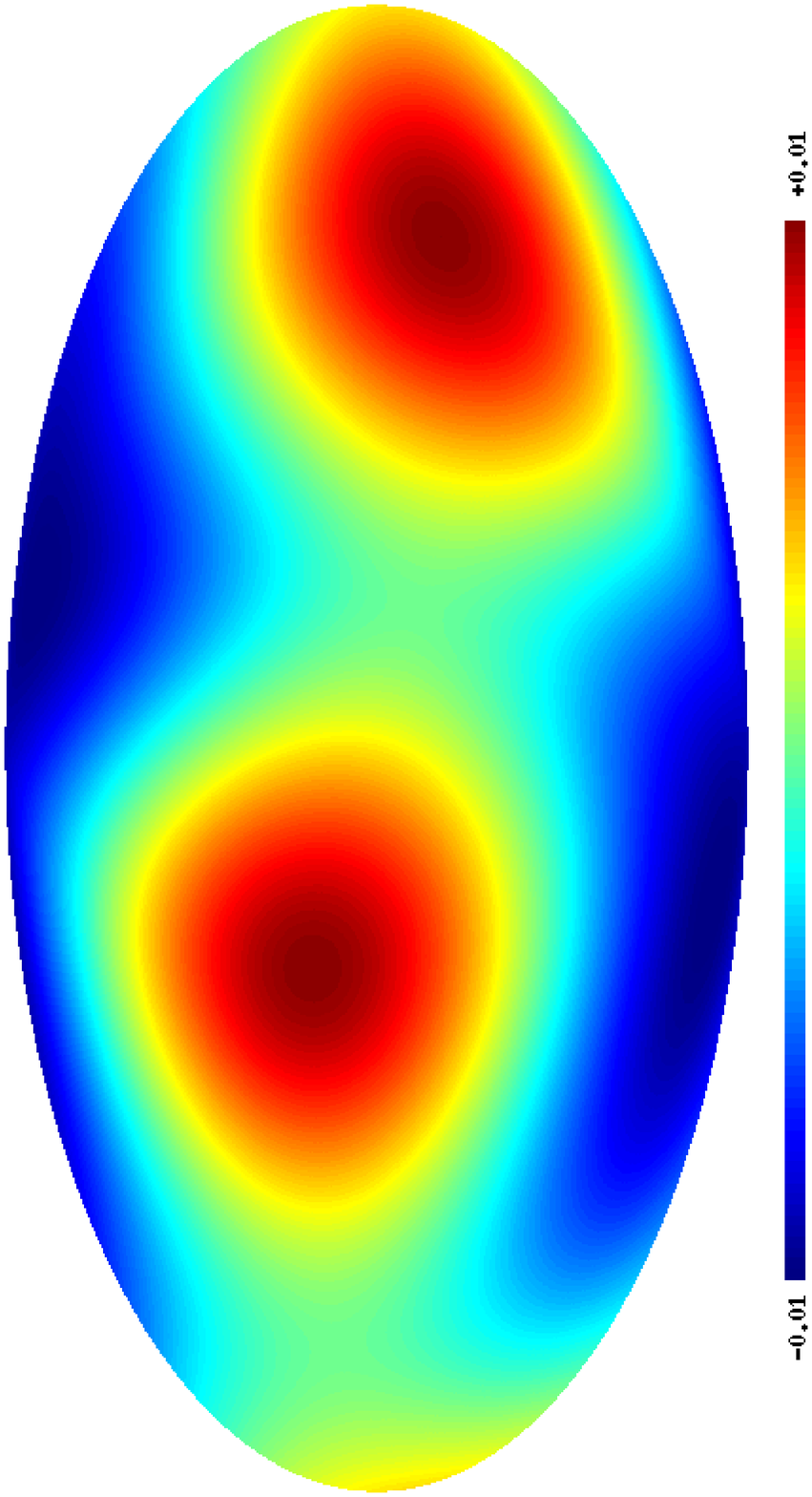,width=6cm}
\caption{
The maps and quadrupoles of the Bianchi VIIh model for
the WMAP\,III.
The top pair is the map for $\l \le 10$ and the quadrupole
%in Ref.~\refcite{jaffe_2006a}.   % 39
The second pair is those
%in Ref.~\refcite{bridges}.       % 25
}
\label{bian}
\end{figure}

As is mentioned in Introduction, starting from the COBE experiment,
the properties of the CMB quadrupole and particularly, the deficit of
its power in comparison with the \wmap
best-fit $\Lambda$CDM cosmological model has attraction of serious attention.
Described in the previous section, phase correlation between the ILC and
foregrounds contain important information beyond the power
spectrum, providing significant restrictions on modification of the theories
of inflation. From our analysis it is clear that models of inflation based
on primordial Gaussian fluctuations of the inflaton field can not explain
the phase correlations. Moreover, the question is if
these cross-correlations can be explained in the framework of
widely discussed the Bianchi VIIh anisotropic cosmological model
(see, e.g.
Refs.~\refcite{jaffe_2006},   % 24
\refcite{jaffe_2006a}         % 39
\refcite{bridges})            % 25
as this model predicts specific
properties of the phases for the CMB quadrupole~? To answer this question
we extract the quadrupoles of the maps derived in
Refs.~\refcite{jaffe_2006a} and \refcite{bridges}.
In Fig.\ref{bian}
we plot both CMB maps for the range of multipoles $\l \le 10$ and
their corresponding quadrupole components. It can be seen clearly that
morphology of these quadrupoles is different from that of
the ILC(III) quadrupole (see Fig.\ref{figD}).

\begin{figure}
\centering
\epsfig{file=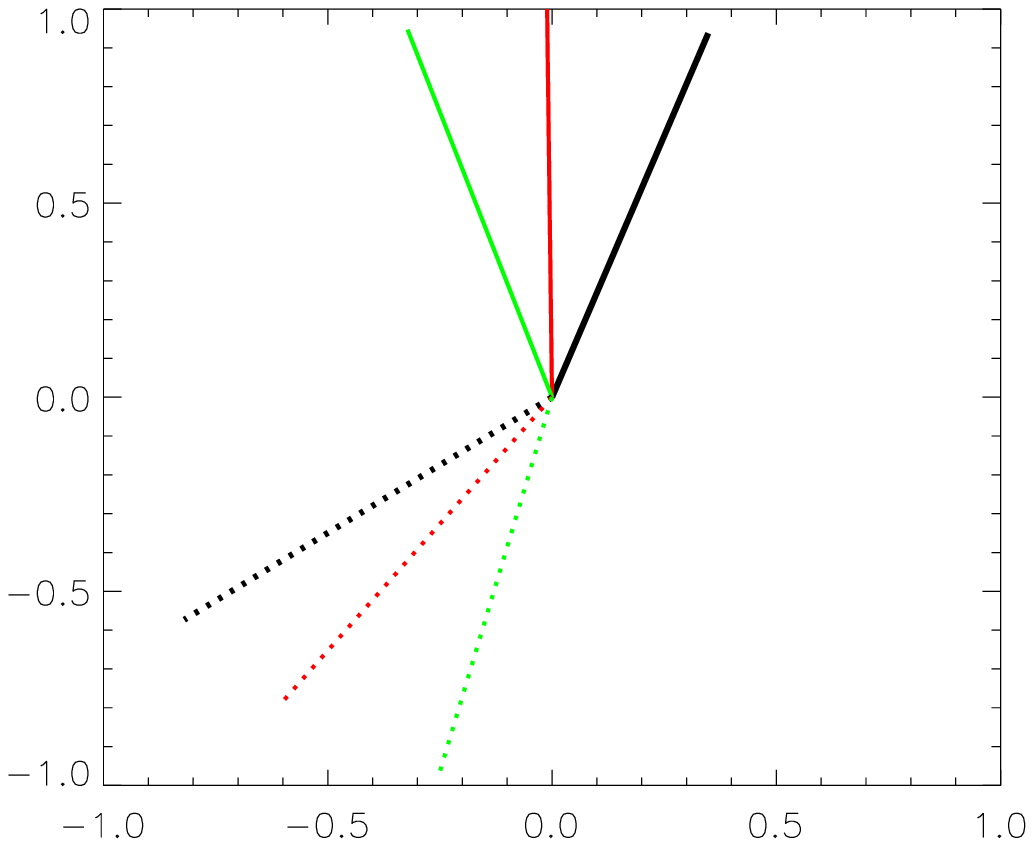,width=6cm}
\caption{
Phases of the quadrupole modes of the ILC(III) and
the Bianchi VIIh model. The phase is represented by the angle subtended
between the positive $x$ axis and the line. Solid lines are for
$(\l,m)=(2,1)$ modes and dotted lines for  $(2,2)$.
Red lines are for ILC(III), black for
Jaffe et al.
% Ref.~\refcite{jaffe_2006a}
and green for
Bridges et al.
% Ref.~\refcite{bridges}.
}
\label{ph}
\end{figure}

To characterize this difference, in Fig.\ref{ph} we plot
the phase diagram for the quadrupole modes of the ILC(III) and
those from the above maps.

In spite of differences in phases between the ILC(III) and both Bianchi VIIh
quadrupoles, they have remarkable correlations. Namely, the phase difference
for the ILC(III) phases is $\Delta_{III}=\xi_{2,2}-\xi_{2.1}\simeq 2.477$
rad, while for
Ref.~\refcite{jaffe_2006a}
quadrupole it is
$\Delta_J=\xi^J_{2,2}-\xi^J_{2.1}\simeq 2.537$ radians and
for Bridges et al. (2006) quadrupole
it is $\Delta_B=\xi^B_{2,2}-\xi^B_{2.1}\simeq 2.560$ radians.
As one can see, with error about $0.07$ radians these angles
$\Delta_{\rm III}, \Delta_J$
and $\Delta_B$ are the same, which means that by simple rotation of
the phases we can get the same morphology of the maps. In particular,
by rotation of the
Ref.~\refcite{jaffe_2006a}
quadrupole phases by the angle
$\beta_J= 0.367$ radians counter-clockwise and by rotation of
Bridges
et al. (Ref.~\refcite{bridges})
phases
by the angle $\beta_B=0.317$ radians clockwise
we get the ILC(III) phases with above-mentioned accuracy.
\begin{figure}
\centering
\epsfig{file=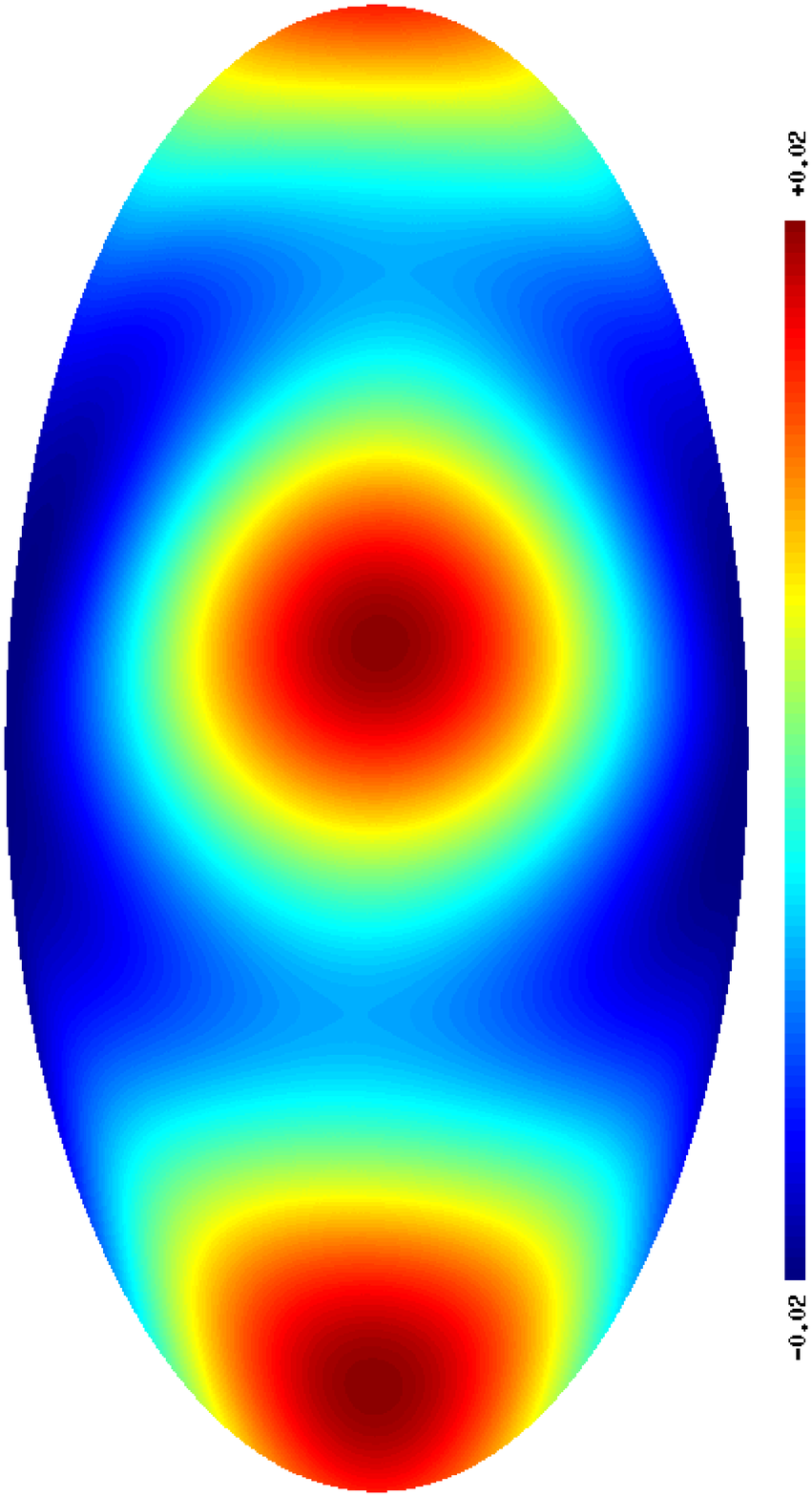,width=6cm}
\caption{
The map for difference between corrected by the angle $\beta_B$
Bridges et al.
% (Ref.~\refcite{bridges})
quadrupole and the ILC(III).
}
\label{bian2}
\end{figure}

In Fig.\ref{bian2} we plot the difference between the Bridges et al.
quadrupole (Ref.~\refcite{bridges})
after rotation of phase by the angle $\beta_B=0.317$ radians
and the ILC(III) quadrupole. Note that the peak to peak amplitude of
the maps shown in Fig.\ref{bian2} are at the same range as the ILC(III)
quadrupole: at $-200,200 \mu$K . That means that in spite of correction
of the phases for Bridges et al.
quadrupole (Ref.~\refcite{bridges}),
the amplitudes of $a_{2,0} \div a_{2,2}$ modes still are not
optimal in comparison with the ILC(III) amplitudes.

%%%%%%%%%%%%%%%%%%%%%%%%%%%%%%%%%%%%%%%%%%%%%%%%%%%%%%%%%%%%%%%%%%%%%%%%%%%%%%%%%%%%%%%%%%

\section{Conclusion}

We have presented the phase analysis of the ILC(I), ILC(III) and de
Oliveira-Costa and Tegmark (Ref.~\refcite{doc_teg})
quadrupoles
in comparison with the foreground phases.
We have shown that the ILC(III) quadrupole has strong correlation with
the foreground phases, mainly with the synchrotron emission per K and KA
band foregrounds.
We have checked out the possibility that the WMAP third year data release
quadrupole can be explained by implementation of the Biachi VIIh model
discussed in Refs.~\refcite{jaffe_2006a} and \refcite{bridges}.
All these models
need additional corrections of the phases for $a_{2,1}$ and $a_{2,2}$
components in order to math the phases of the ILC(III). By analysis of the
de Oliveira-Costa and Tegmark (Ref.~\refcite{doc_teg})
quadrupoles  related with a different
Galactic mask we can conclude that detected above correlations between
the ILC and foregrounds weakly depend on the type of the mask, if the area
covered by them is small in comparison with the $4\pi$.
We have shown that transition from the ILC(I) quadrupole to the ILC(III)
changes significantly the properties of the CMB-foreground
phase correlations
through renormalization of the gain factor. We believe that this is the main
reason for discussed above phase correlation, clearly demonstrated
importance of accurate removal of the systematic effects.

\section*{Acknowledgments}
We thank L.-Y.Chiang, Peter Coles and Carlo Burigana  for useful discussions.
We acknowledge  the use of the Legacy Archive for Microwave Background
Data Analysis (LAMBDA). We also
acknowledge the use of H{\sc ealpix} package
(Ref.~\refcite{healpix})                        % 40
to produce $\alm$ and C.Copi, D. Huterer, D. Schwarz and G.Starkman
for assistance in use their multipole vector package and data.
OVV thanks RFBR for partly supporting by the grant No\,05-07-90139.
The GLESP package (Ref.~\refcite{glesp})        % 41
was used in this work.

\newcommand{\autetal}[2]{{#1\ #2. \etal,}}
\newcommand{\aut}[2]{{#1\ #2.,}}
\newcommand{\saut}[2]{{#1\ #2.,}}
\newcommand{\laut}[2]{{#1\ #2.,}}
%\newcommand{\autetal}[2]{{#1\ #2. \etal,}}
%\newcommand{\aut}[2]{{#1\ #2.,}}
%\newcommand{\saut}[2]{{#1\ #2.,}}
%\newcommand{\laut}[2]{{#1\ #2.,}}

%
% for papers
%
% reference for papers: 1title, 2journal, 3vol, 4page, 5year, 6astro-ph
\newcommand{\refs}[6]{#5, #2, #3, #4} %mnras
\newcommand{\unrefs}[6]{#5, #2 #3 #4 (#6)}  %mnras
%\newcommand{\midunrefs}[5]{#2, #3, #4 (#5);} %prd and prl

%
% for books and proceedings
%
% reference for books: 1title, 2press, 3edition, 4editor, 5year,
% 6astro-ph
%EXAMPLE:
%\book{Numerical Recipes in Fortran}
%{\cup}
%{2nd Ed.}
%{}
%{1992}
%{}
%%\newcommand{\book}[6]{#1, (#2\ #5) } %prd
\newcommand{\book}[6]{#5, {\it #1}, #2} %mnras
%
% reference for proceeding: 1title, 2press, 3editors, 4conf, 5year,
% 6astro-ph
\newcommand{\proceeding}[6]{#5, in #3, #4, #2} %mnras
%EXAMPLE:
%\proceeding{Analysis issues for large CMB data sets}
%{PrintPartners Ipskamp, NL}
%{A. J. Banday, R. S. Sheth and L. Da Costa}
%{Proceedings of the MPA/ESO Cosmology Conference
%``Evolution of Large-Scale Structure''}
%{1999}
%{astro-ph/9812350}
%
%

%\newcommand{\combib}[2]{\bibitem[#1]{#2}} %prd
\newcommand{\combib}[3]{\bibitem[\protect\citename{#1 }#2]{#3}} %mnras

%
% for MNRAS
%
\def\apj{ApJ}
\def\apjl{ApJL}
\def\apjs{ApJS}
\def\mn{MNRAS}
\def\nature{nat}
\def\aa{A\&A}
\def\prl{Phys.\ Rev.\ Lett.}
\def\prd{Phys.\ Rev.\ D}
\def\pr{Phys.\ Rep.}
\def\ijmpd{Int. J. Mod. Phys. D}

\def\cup{Cambridge University Press, Cambridge, UK}
\def\princetonpress{Princeton University Press}
\def\worldpress{World Scientific, Singapore}
\def\oxfordpress{Oxford University Press}

\end{document}